\documentclass{article}
\usepackage{spconf,amsmath,graphicx}
\usepackage{txfonts}
\usepackage{stfloats}
\usepackage{amsmath,graphicx,mathtools,mathrsfs,amssymb}
\usepackage{booktabs}
\usepackage{bm}

\title{Direct source and early reflections localization using deep deconvolution network under reverberant environment}
\name{Shan Gao, Xihong Wu, Tianshu Qu}
\address{Key Laboratory on Machine Perception(Ministry of Education), Speech and Hearing\\ Research Center Peking, University, Beijing, China, qutianshu@pku.edu.com}
\begin{document}
%
\maketitle
\begin{abstract}
This paper proposes a deconvolution-based network (DCNN) model for DOA estimation of direct source and early reflections under reverberant scenarios. Considering that the first-order reflections of the sound source also contain spatial directivity like the direct source, we treat both of them as the sources in the learning process. We use the covariance matrix of high order Ambisonics (HOA) signals in time domain as the input feature of the network, which is concise while contains precise spatial information under reverberant scenarios. Besides, we use the deconvolution-based network for the spatial pseudo-spectrum (SPS) reconstruction in the 2D polar space, based on which the spatial relationship between elevation and azimuth can be depicted. We have carried out a series of experiments based on simulated and measured data under different reverberant scenarios, which prove the robustness and accuracy of the proposed DCNN model.
\end{abstract}
\section{Introduction}
\label{sec:intro}

Direction of arrival (DOA) estimation aim at founding the direction of target sources. It have been implemented in many applications, such as robots \cite{1-2016Robust}\cite{2valin2007robust}, speech enhancement\cite{3gannot2001signal}, room geometry inference\cite{4-2013Room} and so on. With the development of signal process techniques, the DOA methods have attracted wide attention and made great progress in the past few decades. The traditional DOA estimation can be realized based on time difference of arrival (TDOA)\cite{5-georgiou1997robust}, the steered response power (SRP)\cite{6-2000Aibiase} or the subspace algorithms\cite{7schmidt1986multiple}. Since the DOA estimetion methods based on microphone array signals are easily disturbed by coherent signals, many localization algorithms have been proposed in Eigen beam (EB) domain\cite{8bertet20063d}, such as EB-MVDR\cite{9sun2012localization}, EB-MUSIC\cite{10li2011spherical}. The DOA estimation methods in EB domain can estimate accurate spatial feature of sound sources by using frequency smooth algorithm and are more suitable for reverberant environment. Apart from the spatial features, finding a robust and high-resolution mapping between the spatial feature and the source location is also the key to the localization methods. With the development of machine learning algorithms, researchers have proposed many deep neural network (DNN) models for DOA estimation. In contrast to conventional signal processing techniques, neural network-based sound source localization methods require fewer strong assumptions about the environment but obtain the DOA estimation capability with learning process\cite{11chakrabarty2017broadband}\cite{14he2018deep}\cite{15takeda2016sound}\cite{16adavanne2018direction}. These models are shown to perform equally or better than the traditional methods while being robust towards reverberation and noisy scenarios. 

However, such methods also have their problems. One notable point is that there is not a suitable spatial feature as the network input. The input feature like the magnitude and phase component\cite{17adavanne2018sound} will make the network model too complicated, which is not conducive to the improvement of network generalization. The feature like GCC or TDOA\cite{12xiao2015learning}\cite{13vesperini2016neural} will limits the network performance under reverberant environments. Besides, according to the room image source theory\cite{18allen1979image}, both the direct source and early reflections can be regarded as the sources from different directions. Therefore those works that regard the early reflections as distortion in the training process is inappropriate when dealing with reverberation problems.  

A deconvolution network (DCNN) for the direct source and first-order reflections localization under reverberant scenarios is proposed in this paper. We use the covariance matrix of the HOA signals in the time domain as the input feature, which is concise while contains precise spatial information under reverberant environment. Besides, the spatial pseudo-spectrum (SPS) in 2D polar space is reconstructed using deconvolution architecture, based on which the spatial relationship between elevation and azimuth of sources can be 
depicted. 

In the sequel, the paper is structured as follows. The DOA estimation methods based on the covariance matrix in the EB domain is briefly introduced in section 2. The proposed DOA estimation network is described in section 3. The presented technique is experimentally verified using both measured and simulated data in Sec. 4, followed by concluding remarks in Sec.5.

\section{FUNDAMENTIAL THEORY}
\label{sec:pagestyle}

Assuming that the sound field is recorded using a spherical microphone array and transformed into EB domain based on the sound filed decomposition theory\cite{8bertet20063d} , the output HOA signals can be expressed as
\begin{equation}
	\mathbf{B}(kr) = [B_0^0(kr),B_1^0(kr),\dots,B_n^n(kr)]^T,
\end{equation}
where \(B_n^m()\) is HOA signal of order \(n\) degree \(m\). \(k\) is the wavenumber and \(r\) is the radial of the microphone array for recording. The covariance matrix of HOA signals can be calculated as
\begin{equation}
	\mathbf{R}(kr) = \mathbf{B}(kr)\mathbf{B}^H(kr)\ ,
\end{equation}
According to EB-MVDR algorithm, the spatial spectrum of direction \(\Omega_l\) can be estimated as 
\begin{equation}
	P_{EB-MVDR}(kr,\Omega_l) = \frac{1}{\mathbf{Y}^H(\Omega_l)\mathbf{R}^{-1}(kr)\mathbf{Y}(\Omega_l)}\ ,
\end{equation}
\(\mathbf{Y}(\Omega_l)\) is the frequency-independent manifold vector that can be expressed as
\begin{equation}
	\mathbf{Y}(\Omega_l) = [Y_0^0(\Omega_l),Y_1^0(\Omega_l),\dots,Y_n^n(\Omega_l)],
\end{equation}
where \(Y_n^m()\) is the spherical harmonic function of order n degree m. Besides, the subspace-based EB-MUSIC algorithm with high resolution can also be calculated based on the eigenvalue decomposition of the covariance matrix
\begin{equation}
	P_{EB-MUSIC}(kr,\Omega_l) = \frac{1}{\mathbf{Y}^H(\Omega_l)\mathbf{U}_N\mathbf{U}_N^H\mathbf{Y}(\Omega_l)}\ ,
\end{equation}
where the column of the matrix \(U_N\) are the eigenvectors of matrix \(R(kr)\) associated with the \(D\) smallest eigenvalues, and \(D\) is the sources' number. Actually, the EB-MUSIC and EB-MVDR can be regarded as the noise-subspace-based and signal-subspace-based DOA methods, separately. 
The effectiveness of the localization methods mentioned above is based on the accurate estimation of the covariance matrix. To improve the performance of narrowband localization algorithms in reverberant scenarios, the frequency smoothing process is used for broadband-signals cases\cite{9sun2012localization}. Since the manifold vector is frequency independent, the frequency-smoothed covariance matrix can be written as
\begin{equation}
	\tilde{\mathbf{R}}=\sum_{i=1}^{I}\mathbf{R}(k_ir)
\end{equation}
where \(k\in[k_1,k_I]\),\(k_1\) and \(k_I\)denote the lower-end and upper-end frequency bounds of the observed signals. Actually, we can directly obtain the frequency smoothed result by calculating the covariance matrix of  broadband HOA signals in the time domain (\(\mathbf{B}_t\)), which can be denoted as \(\tilde{\mathbf{R}_t}\).

\section{PROPOSED METHOD}
Although the frequency smoothing process could alleviate the ill-conditioning problem of covariance matrix in most scenarios, the spatial resolution and detection sensitivity still limit the EB-MVDR algorithm's performance under high reverberation scenarios. Therefore we propose a deconvolution network-based DOA estimation model to solve this problem. Inspired by the high-resolution EB-MVDR method, we use \(\tilde{\mathbf{R}_t}\) as the input feature of the network, which has two advantages. One is that \(\tilde{\mathbf{R}_t}\) accurately contains spatial information under reverberant environment, the other is that \(\tilde{\mathbf{R}_t}\) is more concise compared with the input features like microphone signals or magnitudes, phases of the spectrograms, which is conducive for the network model learning process.
\label{sec:typestyle}
\subsection{DECONVOLUTION NETWORK}
\begin{figure}[h]
\centering
\includegraphics[width=6cm]{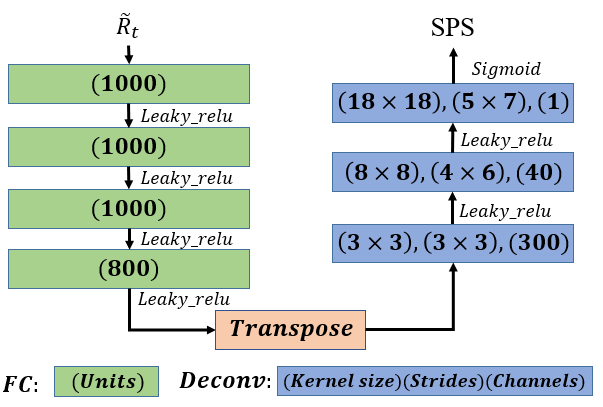}
\caption{Architecture of DCNN.}
\label{DCNN}
\end{figure}
The block diagram of the proposed DCNN network is presented in Fig.\ref{DCNN}. Assuming the order of the Ambisonics signals is N, then the input dimension if the network is \(1\times(N+1)^2\). We use four layers fully connected (FC) network for the spatial feature extracting from \(\tilde{\mathbf{R}_t}\). Then we transpose the feature dimension of the FC layer output to the channel dimension of the following deconvolution (Deconv) layer. We apply a three layers Deconv network for the spatial pseudo-spectrum reconstruction. The output layer that predicts the SPS uses sigmoid activation to estimate multiple directions for a given input feature. Each node of the output layer represents a direction in 2D polar space. The angular resolution of each dimension is 3°, so the output dimension is \(60\times120\). In order to reflect the continuous variation characteristics of the sound field, we perform Gaussian smoothing on the original 0-1 SPS, and the window function can be expressed as
\begin{equation}
	G(\theta, \phi)=\frac{1}{2\pi\sigma^2}e^{-\frac{\theta^2+\phi^2}{2\sigma^2}}
\end{equation}
where \(\theta\) is azimuth and \(\phi\) is elevation. The variance is set to 5, which is obtained using trial method. During testing, the nodes' output is determined with a threshold value of 0.5,  and any great value suggests the presence of a source in the corresponding direction or otherwise absence of sources. We use the cross-entropy of the network output and supervision as the loss function in the training process.


\section{EVALUATION EXPERIMENTS}
\label{sec:typestyle}
To evaluate and compare the performance of the proposed DCNN model, we have carried out a series of experiments on the simulated data and  the measured data under the reverberant environments. We choose the EB-MVDR algorithm as the baseline to verify the effectiveness of the proposed DCNN model. Besides,in order to prove the validity of the proposed input feature of the DNN, we also compare the result of DOAnet[16] with our work. In particular, we use the fourth-oder HOA signals as the input of DOAnet and the same supervise as the DCNN. We use the the mean value \(E_{mean}\) and variance \(E_{var}\) of angle error, precision \(R_{acc}\)  and recall \(R_{rec}\) as the evaluation indicator. Here we set that the DOA estimation with an error of less than 25° is an effectively estimated result, which is about half the width of the beamforming output's main lobe in EB domain\cite{19rafaely2010spherical}. The \(E_{doa}\) is calculated by averaging all angle errors of accurate estimated results, including direct source and first-order reflections. 

\subsection{DATABASE}
For the training and testing of the proposed network, we create a simulation database under different reverberant scenarios based on the image-source model[18]. The room reverberation time is randomly chosen in the range from 300ms to 1000ms. The length, width and height of rectangular rooms range from \(3m\times3m\times2m\) to \(10m\times10m\times4m\). The sources' number is up to 2. The sound field is recorded using a spherical microphone array and transformed into the EB domain up to \(4^{th}\) order. We use the speech signals from the LibriSpeech database as the sources with a sampling rate of 16kHz. The frame size for the calculation of \(\tilde{\mathbf{R}_t}\) is set to 5000 points. We have generated a total of 10000 frames, 80\% for training and 20\% for testing.

\subsection{EXPERIMENTS ON SIMULATED DATA}
\begin{figure}[h]
\centering
\includegraphics[width=8cm]{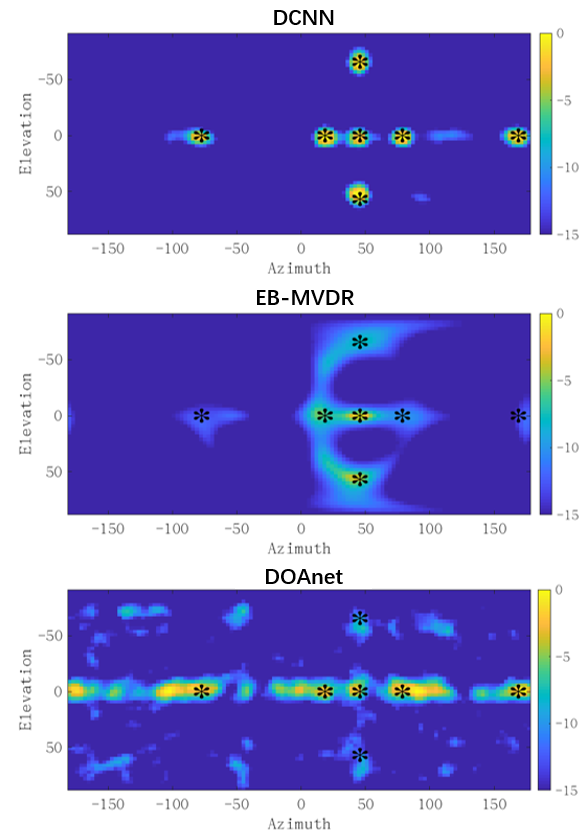}
\caption{SPS of different methods(simulated data).}
\label{simu_map_vs3}
\end{figure}
Fig.\ref{simu_map_vs3} shows the SPS estimation results of EB-MVDR, DOAnet and our proposed DCNN under a rectangle room of size \(4m\times5m\times2.6m\) with \(T_{60}\) as 800ms. Set the lower-left corner of the room as the original point, the coordinate of the source and the microphone array is \((3.0,3.0,1.5)\) and \((2.0,2.0,1.5)\) in meters, separately. In all figures depicting acoustic SPS, the ground truth DOAs for the direct source and first-order reflections are denoted with asterisks. It can be seen that the reflection peaks in SPS of EB-MVDR output is relatively ambiguous compared with DCNN output, which denotes the effectiveness of the proposed network. Besides, it is difficult to distinguish the reflections' direction from the SPS of DOAnet output, which proves the validity of the proposed input feature in the target of reflections localization. In the following experiments, we only make statistics and comparisons on the results of EB-MVDR and proposed DCNN. 

\begin{figure}[h]
\centering
\includegraphics[width=8cm]{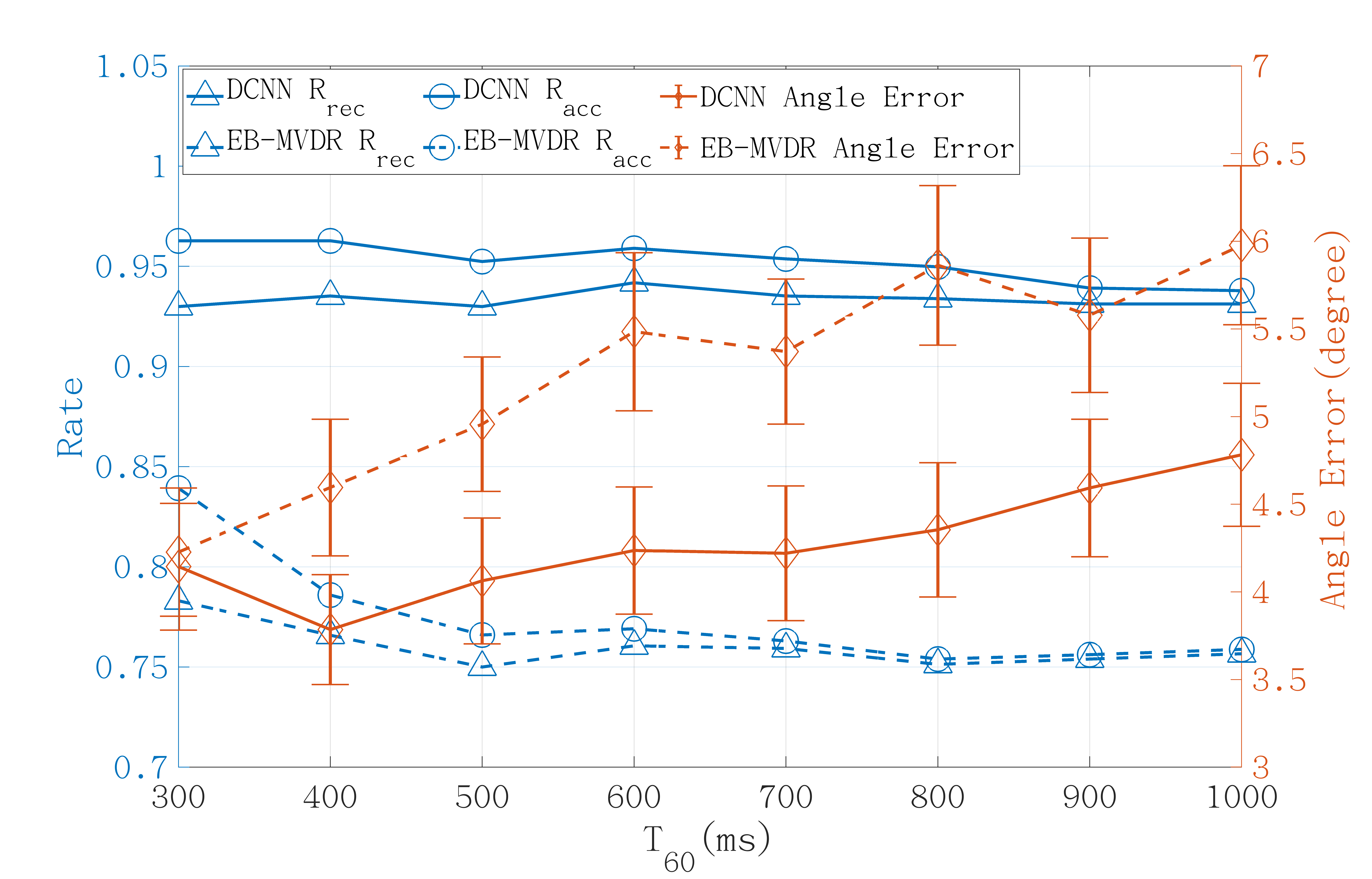}
\caption{DOA results under different reverberant scenarios(one source)}
\label{simu-loss-1s}
\end{figure}

\begin{figure}[h]
\centering
\includegraphics[width=8cm]{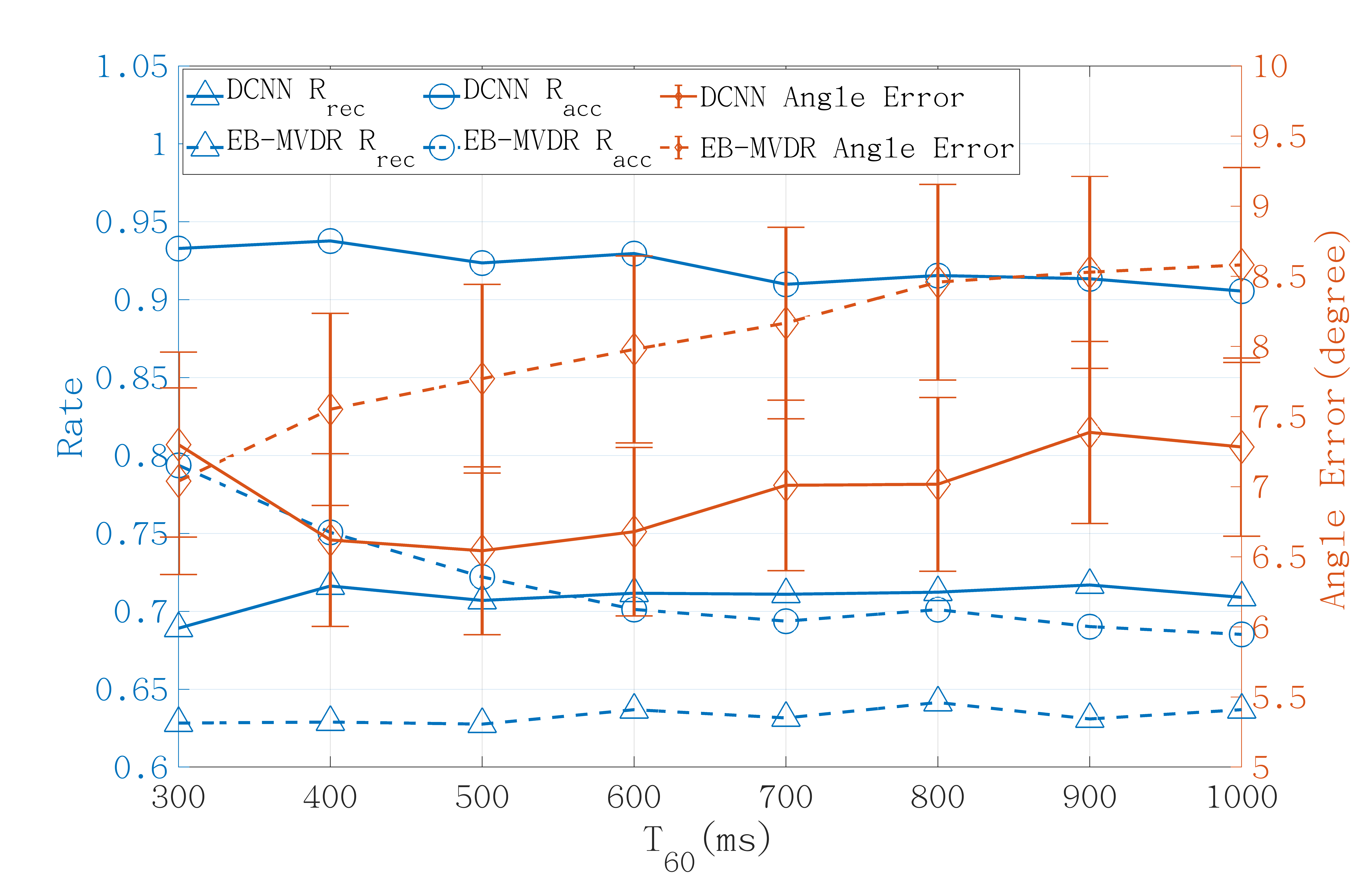}
\caption{DOA results under different reverberant scenarios(two source)}
\label{simu-loss-2s}
\end{figure}

To verify the robustness of the proposed model under different reverberant scenarios, we make statistics on the DOA results under different \(T_{60}\), as shown in Fig.\ref{simu-loss-1s} and \ref{simu-loss-2s}. The blue lines denote the precision and recall, and the orange lines describe the mean and standard deviation of the angle error of different methods. For convenience, the standard deviation range shown in the figure is one-tenth of the actual results, not affecting the relative relationship. It can be seen that the network model method is more stable in different reverberation environments. Besides, the output results of the DCNN network are better than those of EB-MVDR in all circumstances, which shows that the network model can further reduce the interference of coherent signals, and obtain higher stability and accuracy results. It should be pointed that in the two source cases, the reflections of different sources are more likely to overlap or be too close, resulting in the reduction of recall and the increase of error.

\subsection{EXPERIMENTS ON  MEASURED DATA}
In order to evaluate the generalization of the proposed model, we evaluate the performance of our proposed DCNN model using signals measured in a rectangle room with a wood floor, a porous gypsum ceiling, and four smoothed lime walls. The size of the room is 3.38m×5.20m×2.34m. Setting the lower-left corner of the room as the original point, the coordinate of the microphone array is (2.86,1.80, 1.40) in meters. We have placed two sources in the room at (4.27,3.21,1.40) and (1.45,3.47,1.40) in meters, separately. The signals are recorded using the Eigenmike microphone array with 32 elements, allowing us to get the HOA signals up to \(4^{th}\) order. The sources signals, sampling rate and frame length are the same as the previous settings.

\begin{figure}[h]
\centering
\includegraphics[width=8cm]{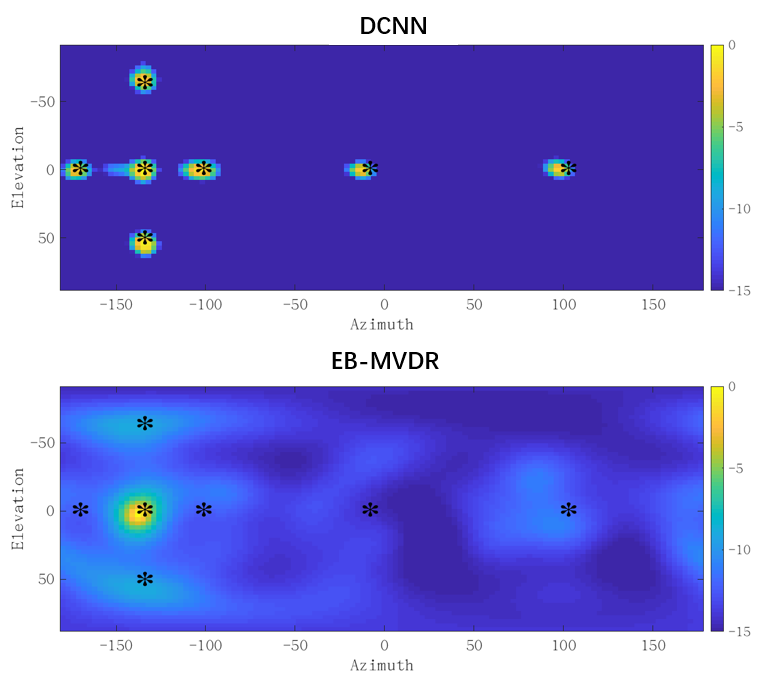}
\caption{SPS of different method(measured data).}
\label{real-1s}
\end{figure}

Fig.\ref{real-1s} shows the estimated SPS of different methods in one source case. It can be seen that the DCNN model trained with simulated data is also effective in the real environment. Compared with the EB-MVDR result, the SPS estimated by the DCNN have better directivity and spatial resolution.  The statistical results of single and double sources cases are shown in Table.\ref{rec_result}. It can be seen that the DCNN model can obtain much accurate DOA estimation results in both cases. Although the precision of DCNN output is a little smaller than that of EB-MVDR result, it can detect reflections more effectively and has a higher recall rate, which shows the effectiveness of the proposed method. 

\newcommand{\tabincell}[2]{\begin{tabular}{@{}#1@{}}#2\end{tabular}} 
\begin{table}[ht]
\centering
\caption{DOA result of measured data}
\label{rec_result}
\setlength{\tabcolsep}{2.1mm}{
\begin{tabular}{cccccc}
\toprule[2pt]
 &	Methods & \(R_{rec}\)& \(R_{acc}\) &\(E_{mean}(^\circ)\)&\(E_{var}(^\circ)\) \\
 \midrule[1pt]
  1s & DCNN & \textbf{0.87} & \textbf{0.88}& \textbf{7.83} & \textbf{6.49} \\
  \midrule[1pt]
  1s & EB-MVDR&0.68 & 0.78& 8.91& 6.77 \\
  \midrule[1pt]
  2s & DCNN& \textbf{0.63}& 0.86& \textbf{8.68}& \textbf{6.05}\\
  \midrule[1pt]
  2s& EB-MVDR& 0.50& \textbf{0.89}& 10.50& 6.79	\\
 \bottomrule[2pt]
\end{tabular}}
\end{table}

\section{CONCLUSION}
We proposed a deconvolutional-based network for DOA estimation of direct source and first-order reflections under reverberant scenarios. We use the covariance matrix of HOA signals in the time domain as the input feature, which even contains the spatial characters of the coherent source. We use the fully connected architecture for the spatial feature extraction and deconvolution network for the SPS reconstruction. The simulated experiments prove the effectiveness and accuracy of the proposed method under different reverberant scenarios compared with the EB-MVDR and DOAnet. Besides, the DCNN model also has better accuracy and effectiveness in recording data, verifying the generalization of the network model trained based on a simulated database.

\bibliographystyle{IEEEbib}

\end{document}